\documentclass[10pt,conference]{IEEEtran}

\usepackage[T1]{fontenc}
\usepackage[utf8]{inputenc}

\usepackage{graphicx}
\usepackage[shortlabels]{enumitem}
\usepackage{amsmath}
\usepackage{amssymb}
\usepackage{amsfonts}

\usepackage{hyperref}  
\hypersetup{pdfborder={0 0 0},colorlinks=true,urlcolor=blue,linkcolor=blue,citecolor=blue,bookmarks=true}
\hypersetup{pdftitle={Detection of Anomalous Network Nodes via Hierarchical Prediction and Extreme Value Theory}}
\hypersetup{pdfauthor={Sevvandi Kandanaarachchi, Mahdi Abolghasemi, Hideya Ochiai, Asha Rao, Conrad Sanderson}}
\usepackage[labelfont={small,bf},textfont=small]{caption}  

\usepackage[british]{babel}
\usepackage{hyphenat}
\usepackage{booktabs}
\usepackage[misc]{ifsym}

\usepackage{bm}
\usepackage{multirow}

\usepackage{fancyvrb}  
\usepackage{newtxtext}  
\usepackage{newtxmath}  
\usepackage{inconsolata}  
\usepackage{dsfont}

\usepackage{balance}  
 
\usepackage{microtype}
\DisableLigatures[f]{encoding = *, family = * }  
\DisableLigatures[f,t,i,?,!,F,I,J]{encoding = *, family = * }

\begin{document}

\title{Detection of Anomalous Network Nodes via Hierarchical Prediction and Extreme Value Theory}

\author
  {
  Sevvandi Kandanaarachchi\textsuperscript{{\tiny~}1},
  Mahdi Abolghasemi\textsuperscript{{\tiny~}2},
  Hideya Ochiai\textsuperscript{{\tiny~}3},
  Asha Rao\textsuperscript{{\tiny~}4},
  Conrad Sanderson\textsuperscript{{\tiny~}1,5}\\
  ~\\
  \textsuperscript{1}{\tiny~}\mbox{\textit{CSIRO,~Australia;}}~
  \textsuperscript{2}{\tiny~}\mbox{\textit{Queensland University~of~Technology, Australia;}}~
  \textsuperscript{3}{\tiny~}\mbox{\textit{University~of~Tokyo,~Japan;}}\\
  \textsuperscript{4}{\tiny~}\mbox{\textit{RMIT~University,~Australia;}}~
  \textsuperscript{5}{\tiny~}\mbox{\textit{Griffith~University,~Australia}}
  }

\maketitle

\begin{abstract}
Continuously evolving cyber-attacks against industrial networks
reduce the effectiveness of signature-based detection methods.
Once malware has infiltrated a network
(for example, entering via an unsecured device),
it can infect further network nodes and carry out malicious activity.
Infected nodes can exhibit unusual behaviour in their use of Address Resolution Protocol (ARP) calls within the network.
In order to detect such anomalous nodes, we propose a two-stage method:
(i) modelling of ARP call behaviour via hierarchical time series prediction methods,
and
(ii) exploiting Extreme Value Theory (EVT) to robustly detect whether deviations from expected behaviour are anomalous.
EVT is able to handle heavy-tailed distributions which are exhibited by internet traffic.
Empirical evaluations on a real-life dataset containing over 10M ARP calls from 362 nodes
show that the proposed method results in considerably reduced number of false positives,
addressing the problem of alert fatigue commonly reported by security professionals. 
\end{abstract}

\begin{IEEEkeywords}
Anomaly detection, local area networks, time series modelling, hierarchical prediction, extreme value theory, network security.
\end{IEEEkeywords}

\section{Introduction}

\begin{figure}[!b]
\vspace{-3ex}
\centering
\includegraphics[width=0.90\columnwidth]{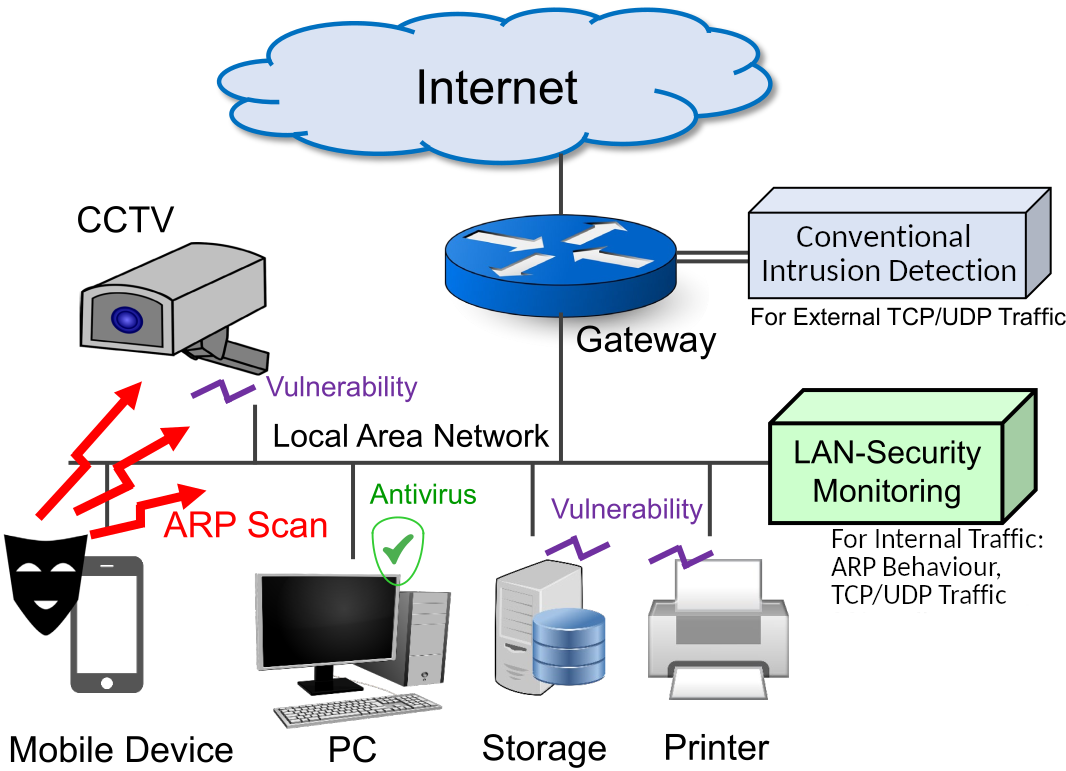}
\vspace{1ex}
\caption
  {
  ARP scan calls made by LAN-internal malware are not visible to conventional network intrusion detection systems, which only observe incoming/outgoing traffic (such as TCP/UDP).
  An internal monitoring device attached to the LAN can observe ARP calls and detect unusual behaviour.
  }
\label{fig:example}
\end{figure}

Approaches for defending industrial networks against cyber-attacks can be generally placed into three main levels:
\textbf{(i)}~at the gateway to the wider internet,
\textbf{(ii)}~at each node,
\textbf{(iii)}~across the local network.
A conceptual demonstration of these approaches is shown in Fig.~\ref{fig:example}.
At level~(i), traffic incoming to the local network is scanned and compared against known patterns of malware and infiltration techniques~\cite{AhmetogluDas22,khraisat2019survey}.
At level~(ii), antivirus software installed on individual nodes performs a vital role in catching known malware.
At level~(iii), internal network traffic is analysed to detect unusual behaviour~\cite{Guo_2025}.

The changing nature and increasing sophistication of attacks makes the detection of intrusion attempts at all levels more difficult~\cite{AhmetogluDas22}.
Detection at levels~(i) and~(ii) may even be entirely bypassed,
as malware can enter the network through an unsecured device.
Furthermore, many false alarms can lead to \textit{alert fatigue},
which in turn reduces the effectiveness of addressing critical security incidents~\cite{Ban2021,Tariq_2025}.
Explicitly targeting a low false positive rate is hence a critical characteristic of practical infiltration detection methods.

In this work we focus on monitoring Address Resolution Protocol (ARP) calls within the network to detect unusual behaviour that can be indicative of malware.
ARP spoofing and poisoning attacks make the protocol a vulnerability in the LAN architecture~\cite{AHUJA2022107757,balogh2018lan,GIRDLER2021106990}.
Rather than comparing against preset patterns of known malware,
we first learn the typical behaviour of network nodes and then detect deviations from the learned behaviour.

More specifically,
we propose to use hierarchical prediction methods which leverage information from multiple nodes in the network to predict node behaviour,
unlike conventional prediction methods that investigate each node in isolation.
Furthermore, in order to explicitly minimise the false positive rate,
the deviations from predicted behaviour are analysed via Extreme Value Theory (EVT)~\cite{Coles_2001,Reiss_2007},
which has been shown to be effective in detecting anomalies within the context of dynamic graphs~\cite{Kandan_2024}.
EVT can handle heavy-tailed distributions that are exhibited by computer network traffic~\cite{HernandezCampos2004},
which does not fade quickly as in a normal (Gaussian) distribution.
Overall, the approach is designed to be interpretable as~it~is~based~on statistical techniques,
in contrast to methods employing \mbox{black-box} algorithms~\cite{Barredo_Arrieta_2020,Sanderson_2023}.
To the best of our knowledge, the combination of hierarchical prediction and EVT is the first time such an approach has been used in computer network security.

We continue the paper as follows.
Section~\ref{sec:proposed_method} describes the proposed method in more detail.
Section~\ref{sec:experiments} provides an empirical evaluation of the proposed method on a real-life dataset containing over 10 million ARP calls from 362 nodes.
The main findings are summarised in Section~\ref{sec:conclusion}.

\newpage

\section{Proposed Method}
\label{sec:proposed_method}
\vspace{-0.20ex}

Hierarchical time series refers to a collection of time series that are nested in a multi-level structure.
In general, series at various levels of the hierarchy can be used to extract latent information from other levels
that may not be apparent in any one single level~\cite{abolghasemi2022model}. 
In this work we consider a two level hierarchy
and use hierarchical time series to predict the number of ARP calls made by each node.

The top level in the hierarchy contains the sum of the time series,
which gives the general pattern of the network.
The bottom level contains the individual nodes of the network.
Time series modelling is used for each node,
followed by leveraging the hierarchical structure of the network to reconcile the predictions for the nodes
(ie., the original predictions are adjusted by taking into account the total ARP calls)~\cite{abolghasemi2022model}.

For reconciliation we use the well-grounded \textit{Minimum Trace} 
technique~\cite{spiliotis2021hierarchical,Wickramasuriya_2019},
briefly summarised as follows.
The~reconciled time series predictions are given by
\mbox{\small $\widetilde{\bm{Y}} = \bm{S} \bm{G} \widehat{\bm{Y}}$},
where {\small $\widehat{\bm{Y}}$} are the base predictions,
while {\small $\bm{G}$} is a mapping matrix and {\small $\bm{S}$} is a summing matrix.
The optimal {\small $\bm{G}$} minimises the trace of the covariance matrix of reconciled prediction errors,
and can be found via
\mbox{\small $\bm{G} = (\bm{S}^T \bm{W}^{\dagger} \bm{S})^{-1} \bm{S}^T \bm{W} ^{\dagger}$},
where {\small $\bm{W}$} is the variance–covariance matrix of base prediction errors,
while {\small $\bm{W}^\dagger$} is the generalised inverse of {\small $\bm{W}$}.
The \textit{shrinkage} approach is used for estimating {\small $\bm{W}$}~\cite{Lancewicki_2014,Ledoit_2022}. 

We consider four time series modelling methods with varying complexity:
Error~Trend~Seasonal~(ETS),
Time Series Linear Regression (TSLM),
Zero-Inflated Negative Binomial (ZINB),
and 
Light Gradient Boosting Model (LightGBM).
ETS is a straightforward but effective statistical method based on state-space exponential smoothing, successfully used in many prediction tasks~\cite{fpp3}.
TSLM regresses the response variable (signals) as a linear function of the explanatory variables.
We use the trend and the last six observations of the series as the explanatory variable.
A logarithm of signals is used as the response variable, enabling the log-linear model to better estimate the non-linear behaviour of the signal time series~\cite{fpp3}.
A small scalar is added to the signals to sidestep problems with taking the logarithm of zero-valued data.
In contrast, ZINB pays considerable attention to the many zeros present in the signal time series data,
while at the same time taking into account the large spikes that can occur after zeros~\cite{Cameron_2005}.
We use the last six observations of signals as the regressors of the model.
LightGBM is a popular gradient boosting framework based on decision trees~\cite{Ke_2017}.
To minimise the computational burden, the hyperparameters of the LightGBM model are optimised via grid search. 

We define an anomaly as a point with low conditional probability.
Conditioned on its previous behaviour, the probability of a node's behaviour at time $t$ is computed using residuals
(differences between predicted and actual behaviour).
In other words, a node is deemed as anomalous if its behaviour is substantially different with respect to its previous behaviour.
This formalisation is suitable for describing heterogeneous behaviour,
where each node may behave in a considerably different manner compared to other nodes.

\begin{figure}[!b]
  \vspace{-1ex}
  \centering
  \footnotesize
  \begin{minipage}{\columnwidth}
  \begin{minipage}{0.05\columnwidth}
    \centering
    \footnotesize
    \textbf{(a)}
  \end{minipage}
  \hfill
  \begin{minipage}{0.90\columnwidth}
    \centering
    \includegraphics[width=\textwidth]{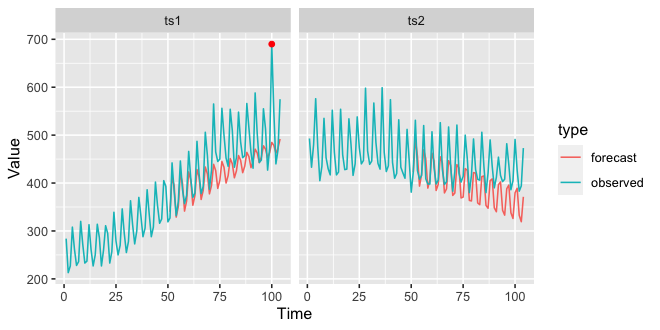}
  \end{minipage}
  \end{minipage}
  \begin{minipage}{\columnwidth}
  \begin{minipage}{0.05\columnwidth}
    \centering
    \footnotesize
    \textbf{(b)}
  \end{minipage}
  \hfill
  \begin{minipage}{0.90\columnwidth}
    \centering
    \includegraphics[width=\textwidth]{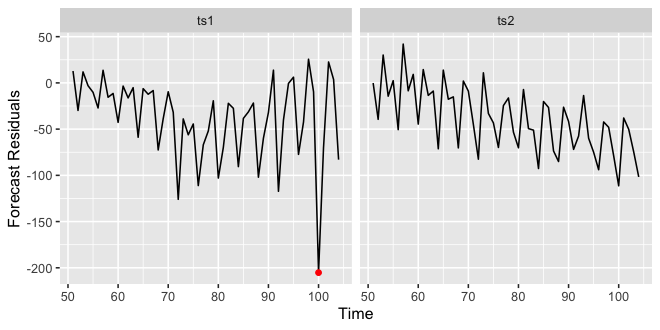}\\
  \end{minipage}
  \end{minipage}
  \vspace{-1ex}
  \caption
    {
    \textbf{(a)} Two example time series (ts1 and ts2), with original data shown in blue and predicted data in red;
    the time span of $t=1$ to $t=50$ was used for training the prediction method.
    \textbf{(b)} Differences (residuals) between actual and predicted values for ts1 and ts2.
    For~ts1, an anomaly is depicted as a red dot at time $t = 100$ in both~(a) and~(b).
    }
  \label{fig:exampletwots}
\end{figure}

As an example, consider two time series of length 104,
shown in Fig.~\ref{fig:exampletwots}(a),
where the first 50 time points for used training and the remainder for testing.
An anomaly at time $t = 100$ in the first time series is denoted by a red dot.
Via time series modelling, the predicted portions of the time series are shown in red.
The differences between actual and predicted time series are shown in Fig.~\ref{fig:exampletwots}(b),
with the anomalous point yielding a large negative residual.

More formally, 
we denote the expected behaviour of node $i$ at time $t$ via prediction $\widehat{y}_{i,t}$
and actual behaviour as $y_{i,t}$.
For well-modelled data, the residuals $e_{i,t} = y_{i,t} - \widehat{y}_{i,t}$ are assumed to be normally distributed,
meaning that the probability of large absolute residuals is assumed to be low.
In turn, anomalous behaviour is assumed to result in large residuals and have low conditional probability,
\mbox{$\vert e_{i,t} \vert \geq c\left( \epsilon \right) \iff  p_{i,t\vert t-1} \leq \epsilon$},
which indicates that anomalies can be detected by analysing the $e_{i,t}$ residuals.

However, real-life data can exhibit complex patterns,
including heavy-tailed distributions as shown by internet traffic~\cite{HernandezCampos2004}.
This indicates that a straightforward classifier based on thresholding the residuals
\mbox{(ie., $\vert e_{i,t} \vert > \text{threshold} \Rightarrow$ anomaly)}
can result in many erroneous classifications,
leading to high false positive rates.

To address the above problem, we use EVT to explicitly handle heavy-tailed distributions~\cite{Coles_2001,Reiss_2007},
as per the implementation in~\cite{Kandan_2022}.
In the context of anomaly detection, EVT has been shown to result in low false positives~\cite{Kandan_2024,Talagala2021}.
The residuals are modelled as a Generalised Pareto Distribution (GPD),
which takes into account extreme values.
The GPD is defined as:

\vspace{-2ex}
\begin{small}
\begin{equation}
   H(v) = 1 - \left( 1 + {\left(\xi v\right)}~/~{\sigma_u} \right)^{-1/\xi}
  \label{eq:POT1}
\end{equation}
\end{small}

\noindent
where the domain of $H$ is {\small $\{v: v >0\, \, \text{and} \, \,  (1 + \xi v/\sigma_u) >0  \}$},
and {\small $\sigma_u = \sigma + \xi(u- \mu)$}.
Parameters $\mu$, $\sigma$ and $\xi$ denote the location, scale and shape of the distribution, respectively.
{\small $\xi \mbox{~=~} 0$} specifies a Gumbel distribution,
{\small $\xi \mbox{~<~} 0$} specifies a Weibull distribution,
and
{\small $\xi \mbox{~>~} 0$} specifies a Fréchet distribution.
Parameter $u$ is used for selecting a subset of the given samples that represent extremes,
and is commonly set to the 90-th percentile~\cite{Kandan_2022}.
The remaining GPD parameters are estimated by numerically maximising a log-likelihood function,
detailed in~\cite{Coles_2001,Reiss_2007}.

EVT guarantees that the tail of all well-behaved distributions is categorised by Gumbel, or Weibull, or Fréchet distributions.
For example, the tail of the normal distribution is described by a Gumbel distribution.
The normal distribution has an exponentially decaying tail,
which is generally considered easy to model adequately if the anomalies actually display such behaviour.
However, a more challenging scenario is when the tail decays according to a power law.
If the tail is not modelled correctly,
large non-anomalous data points can be easily but incorrectly deemed anomalous,
increasing the number of false positives.
In this case, a power law tail is better modelled by a Fréchet distribution,
leading to a decrease in the number of false positives.
The GPD accommodates all tail behaviours,
without separately fitting the Gumbel, Weibull and Fréchet distributions. 
Fitting a GPD is a data-driven approach that accommodates varying tail behaviour
without imposing a fixed distribution on it.

The probability of observing a value higher than $v_i$ is obtained via
{\small $P(v \geq v_i) = \int_{v_i}^{\infty} H(v) ~ dv$}.
A user-configured threshold is then applied on the obtained probability
to classify a given sample as anomalous.
If {\small $P(v \geq v_i) \leq \alpha$},
then $v_i$ is deemed anomalous.
We use $\alpha = 0.05$ in all experiments.

\section{Empirical Evaluation}
\label{sec:experiments}
\vspace{-1ex}

\subsection{Dataset}
\label{sec:dataset}

\begin{figure}[!tb]
  \begin{minipage}{1.0\columnwidth}
  \begin{minipage}{1.0\columnwidth}
    \centering
    \includegraphics[width=\columnwidth,height=0.7\columnwidth]{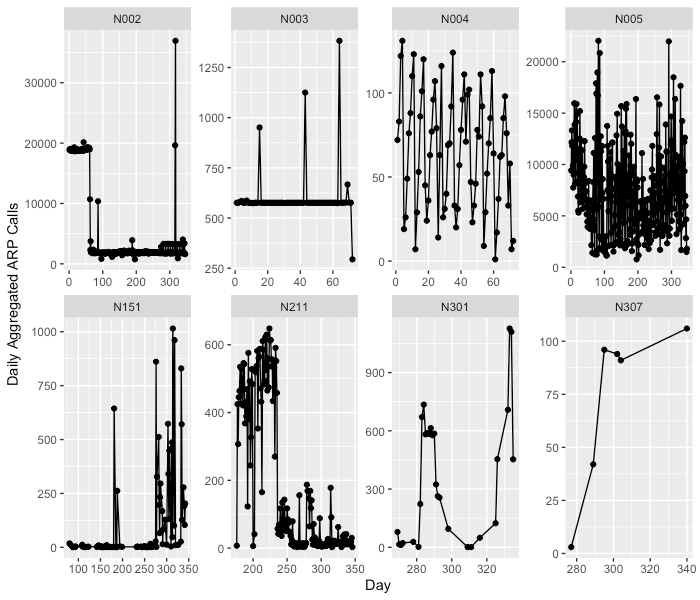}
    \caption
      {
      Examples of variations in daily aggregated ARP call patterns from 8 representative nodes, 
      obtained over approximately one year.
      }
    \label{fig:arpcalls}
  \end{minipage}
  \vspace{3ex}
  \begin{minipage}{1.0\columnwidth}
  \end{minipage}
    \centering
    \includegraphics[width=\columnwidth,height=0.45\textwidth]{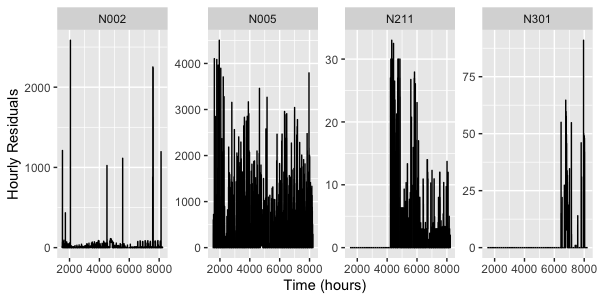}
    \caption
      {
      Hourly residuals, resulting from fitting via ETS.
      }
    \label{fig:residualsbynode} 
  \end{minipage}
\end{figure}

The dataset used in the experiments is comprised of 10,309,621 ARP calls from 362 nodes,
observed over approximately a year.
It was obtained as part of a larger LAN-security monitoring project~\cite{Sun_2021}.
The dataset was generated by deploying a monitoring device on a designated local area network.
The device monitored both the traffic broadcast over the entire network as well as the packets directed at the monitoring device.
The captured broadcast traffic includes ARP requests.
An ARP request has source and target IP addresses for looking up the MAC address for forwarding IP packets.
Each ARP request is identified with the sender's MAC address as the source IP address may change based on the assigning policy of the DHCP server of the network.
Fig.~\ref{fig:arpcalls} contains examples of variations in daily aggregated ARP call patterns,
while Fig.~\ref{fig:residualsbynode} illustrates the considerable variations in the magnitude of the hourly residuals,
resulting from fitting via the ETS model~\cite{fpp3}.

An attack vector originating from a compromised node can occur as follows.
The node first aims to find available hosts in the network by broadcasting IP packets and expecting responses from the hosts,
aiming for discovery of available open TCP/UDP ports for further intrusion and/or exfiltration of data from file servers.
During this activity, many ARP requests with changing target IP fields are observed as the sender wants to know the target MAC addresses.

When such an attack is carried out on a host in the network,
the monitoring device also receives IP packets at the TCP/UDP layer from the sender,
along with the observed sequences of broadcast ARP requests.
Under the assumption that there is no legitimate reason to send a TCP/UDP packet to the monitoring device,
nodes accessing the monitoring device can be considered as anomalous.
In this context, the monitoring device is treated as a source for ground truth labels,
resulting in the dataset containing about 1\% anomalies.

\subsection{Protocol}

The evaluation protocol is first summarised as follows.
The input is a collection of ARP time series for each node in the network,
with a specified time series prediction method, as well as size of training and test windows.
Using data in the training window we compute hierarchical predictions for each node.
Using the test window we compute the prediction residuals $e_{i,t}$ for each node.
We then use EVT to identify anomalies.
Once the anomalies are identified for the given test window,
the training and test windows are advanced to include the newer data. 
We consider aggregate behaviour at hourly intervals.
Further details of the evaluation protocol are elucidated in the following paragraphs.

The original data is of the form $\left(t_i, \text{node}_{t_i}\right)$.
For a given time $t_i$ there can be more than one host, $\text{node}_{t_i}$, making ARP calls in the network.
We transform the data into a time series format by considering hourly aggregates and rearranging these by node id.
Thus, for each node $i$ we obtain a time series $\{y_{i,t} \}_{t = 1} ^N$.
Note that for certain time periods $y_{i,t}$ can be zero.

We consider an initial time period of 8 weeks to train the time series prediction models,
denoting this period by~\mbox{$[t_0, t_{\text{train}_1}]$}.
The trained models are then used to predict hourly ARP calls for each node for the next week.
This forms the first test period,
spanning $[t_{\text{train}_1}, t_{\text{test}_1}]$,
where $t_{\text{test}_1}$ denotes the end of the first test week.
For each week we have $24 \times 7$ predictions for each node.
We use a rolling/expanding window model in which the \mbox{$i$-th} training period spans $[t_0, t_{\text{train}_i}]$
and the \mbox{$i$-th} test period spans the week $[t_{\text{train}_i}, t_{\text{test}_i}]$.
This results in approximately 42 weeks of rolling hourly predictions for each node,
corresponding to 7056 hourly blocks.

After fitting a hierarchical time series model and predicting hourly ARP calls by node for each test week,
the residuals are computed.
The hourly residuals are then fed into the EVT-based classifier which identifies the anomalous ARP calls. 
Fig.~\ref{fig:evaluation_protocol} shows the detailed pseudo-code for the evaluation protocol.

\begin{figure}[!b]
\begin{minipage}{1\columnwidth}
\hrule
\vspace{1ex}
\fontsize{7.5}{9.0}\selectfont
\textbf{Input:}
\begin{enumerate}[{$\bullet$}]
\item collection of ARP time series $\{ y_{i,t} \}_{t = 1}^N$, with $i$ being the node identifier
\item initial training window length $L_{\text{train}_1}$
\item test window length $L_{\text{test}}$
\item frequency of predictions (eg., hourly)
\item time series prediction method $\in$ \{~ETS, TSLM, ZINB, LightGBM~\}
\end{enumerate}

\vspace{1ex}

\textbf{Output:}
\begin{itemize}[{$\bullet$}]
\item anomalous nodes and respective timestamps
\end{itemize}
\vspace{1ex}

\textbf{Evaluation:}
\begin{itemize}[{$\bullet$}]
\item let $t_{0}$ and $t_{\max}$ denote the initial and final time stamps
\item let $j = 1$ and $t_{\text{train}_j} = t_0 + L_{\text{train}_1}$
\item let $t_{\text{test}_j} = t_{\text{train}_j} + L_{\text{test}}$
\item do while $t_{\text{test}} \leq t_{\max}$:
\begin{itemize}[{$\circ$}]
\item form training set $Y_{\text{train}} = \{y_{i,t}\}$ for $t < t_{\text{train}_j}$ for all $i$ in collection
\item form test set $Y_{\text{test}} =  \{y_{i,t}\}$ for $ t_{\text{train}_j} \leq t < t_{\text{test}_j}$ for all $i$ in collection
\item fit time series prediction model to $Y_{\text{train}}$ and obtain initial predictions
\item use a two-level hierarchy and reconcile the initial predictions
\item let $\widehat{y}_{i,t}$ = reconciled individual predictions for each node \& each time stamp
\item let $e_{i,t} = y_{i,t} - \widehat{y}_{i,t}$ denote the residuals
\item use EVT on $\{e_{i,t}\}$ to find anomalies
\item $j := j + 1 $
\item let $t_{\text{train}_j} = t_{\text{train}_j} + L_{\text{test}}$ to advance the training set
\item let $t_{\text{test}_j} = t_{\text{train}_j} + L_{\text{test}}$ to advance the test set
\end{itemize}
\end{itemize}
\hrule
\normalsize
\end{minipage}
\vspace{1ex}
\caption{Evaluation protocol pseudo-code for the proposed method.}
\label{fig:evaluation_protocol}
\end{figure}

\subsection{Evaluation}

Four variants of the proposed approach are evaluated,
with the variants differing in the employed hierarchical prediction method.
All variants use EVT to determine whether the residuals represent an anomaly.
The variants are:
\textbf{(i)}~ETS~+~EVT;
\textbf{(ii)}~TSLM~+~EVT;
\textbf{(iii)}~ZINB~+~EVT.
\textbf{(iv)}~LightGBM~+~EVT;
A~two-level hierarchy is used in all variants, with 362 nodes at the bottom level indicating 362 MAC addresses,
and one node at the top indicating the total ARP calls in the network.

As autoencoders are popular in unsupervised anomaly detection~\cite{provotar2019unsupervised,khamphakdee2014improving},
the four variants are also compared against a popular autoencoder-based network anomaly detection method detailed in~\cite{chen2018autoencoder}.
Generally an autoencoder reconstructs the input via an encoder and a decoder framework.
If the reconstruction error is larger than a certain percentile,
it is identified as anomalous.
To be consistent with the proposed approach,
we used $(1- \alpha)$-th percentile with $\alpha = 0.05$ in all experiments,
with residuals $e_{i,t}$ in the rolling window as input to the autoencoder. 
Other hyper-parameters were kept at their default settings.

Table \ref{tab:results_mint} summarises the results in terms of mean values for precision, recall and F-measure~\cite{Tharwat_2020},
as well as the number of false positives per window.
The results show that LightGBM variant has the highest \mbox{F-measure},
followed by the TSLM and ZINB variants.
The lowest F-measure is achieved by the autoencoder.
The \mbox{LightGBM}, ETS and ZINB variants have very low false positives per each time window,
which is in contrast to the autoencoder that obtains a much higher number of false positives.
The~\mbox{LightGBM} variant obtains considerably better precision than the other techniques.
While the autoencoder achieves the highest recall, it comes at the expense of very low precision.
A higher recall is expected when many points are identified as belonging to the positive class.
In general, the results demonstrate the efficacy the proposed approach,
with the LightGBM variant obtaining best overall performance.

\begin{table}[!tb]
\centering
\caption
  {
  Anomaly detection performance of the autoencoder and four variants of the proposed approach.
  Results are reported in terms of mean value.
  }
\footnotesize
\begin{tabular}{rcccc}
\toprule
\textbf{Method / Variant} & \textbf{Precision} & \textbf{Recall} & \hspace{-0.5ex}\textbf{F-measure}\hspace{-0.5ex} & \textbf{False Pos.}\hspace{-1ex} \\
\midrule
Autoencoder    & 0.018          & \textbf{0.890} & 0.032          & 59.049         \\ \midrule
ETS + EVT      & 0.286          & 0.769          & 0.208          &  7.634         \\ \midrule
TSLM + EVT     & 0.291          & 0.790          & 0.256          & 17.098         \\ \midrule
ZINB + EVT     & 0.318          & 0.735          & 0.243          & 3.341          \\ \midrule
LightGBM + EVT & \textbf{0.585} & 0.659          & \textbf{0.341} & \textbf{1.000} \\
\bottomrule
\end{tabular}  
\label{tab:results_mint}
\end{table}

\section{Concluding Remarks}
\label{sec:conclusion}

In this work we have proposed an approach for improving security in industrial networks
through the detection of anomalous nodes that exhibit behaviour indicative of infection by malware.
The method relies on monitoring all Address Resolution Protocol (ARP) calls throughout a local area network.
ARP call behaviour for each node is modelled via hierarchical time series prediction,
followed by employing Extreme Value Theory (EVT) to robustly determine
whether differences from expected behaviour (residuals) signify anomalies.
Empirical evaluations on a real-life dataset containing over 10M ARP calls in a network comprised of 362 nodes
show that the proposed approach results in a relatively high F-measure and a considerably reduced number of false positives.

Time series prediction methods are able to model complex trend and seasonality patterns in data~\cite{abolghasemi2022model},
which can account for fluctuations in the network traffic with respect to time of day or day of the week. 
In contrast to modelling the behaviour of each node in isolation, 
the hierarchical prediction enables the extraction of more information from multiple nodes to predict the behaviour of a particular node,
leading to more accurate predictions for all nodes.

Residuals allow the examination of deviations from expected behaviour,
while taking into account heterogeneity of behaviours across nodes.
For example, observed values from a node can be both time- and user-dependent.
A node used by a particular user can result in a high volume of network traffic at certain hours.
While this type of behaviour might be expected from this particular node,  
it can be anomalous for other nodes.

Internet traffic can exhibit patterns following heavy-tailed distributions~\cite{HernandezCampos2004}.
EVT explicitly takes into account such distributions,
which leads to low false positive rates in anomaly detection.
This is a critical aspect,
as high false positive rates can cause \textit{alert fatigue} for security professionals~\cite{Ban2021,Tariq_2025},
which in turn risks security lapses
that can lead to exfiltration of sensitive data.

Further avenues of research include the evaluation of alternative hierarchical time series reconciliation methods~\mbox{\cite{abolghasemi2022model,spiliotis2021hierarchical}},
as well as more elaborate hierarchy configurations (eg., \mbox{3-level} hierarchy),
where similar network devices are placed into groups to form an additional level within the hierarchy.

~

\begin{footnotesize}
\noindent
\textbf{Acknowledgements}.
Sevvandi Kandanaarachchi is part of the Australian Research Council (ARC)
Industrial Transformation Training Centre in Optimisation Technologies,
Integrated Methodologies, and Applications (OPTIMA),
Project ID IC200100009.
\end{footnotesize}

\def~{\,}  
\balance

\bibliographystyle{ieee_mod}  
\bibliography{references}

\begin{thebibliography}{10}\interlinepenalty=10000\itemsep=0.4ex

\bibitem{abolghasemi2022model}
M.~Abolghasemi, R.~J. Hyndman, E.~Spiliotis, and C.~Bergmeir.
\newblock Model selection in reconciling hierarchical time series.
\newblock {\em Machine Learning}, 111(2):739--789, 2022.

\bibitem{AhmetogluDas22}
H.~Ahmetoglu and R.~Das.
\newblock A comprehensive review on detection of cyber-attacks: Data sets,
  methods, challenges, and future research directions.
\newblock {\em Internet of Things}, 20:100615, 2022.

\bibitem{AHUJA2022107757}
N.~Ahuja, G.~Singal, D.~Mukhopadhyay, and A.~Nehra.
\newblock Ascertain the efficient machine learning approach to detect different
  {ARP} attacks.
\newblock {\em Computers and Electrical Engineering}, 99:107757, 2022.

\bibitem{balogh2018lan}
Z.~Balogh, {\v{S}}.~Koprda, and J.~Francisti.
\newblock {LAN} security analysis and design.
\newblock In {\em International Conference on Application of Information and
  Communication Technologies (AICT)}, 2018.

\bibitem{Ban2021}
T.~Ban, N.~Samuel, T.~Takahashi, and D.~Inoue.
\newblock {Combat Security Alert Fatigue with AI-Assisted Techniques}.
\newblock In {\em Cyber Security Experimentation and Test Workshop}, pages
  9--16. ACM, 2021.

\bibitem{Barredo_Arrieta_2020}
A.~Barredo~Arrieta et~al.
\newblock Explainable artificial intelligence ({XAI}): Concepts, taxonomies,
  opportunities and challenges toward responsible {AI}.
\newblock {\em Information Fusion}, 58:82--115, 2020.

\bibitem{Cameron_2005}
A.~C. Cameron and P.~K. Trivedi.
\newblock {\em Microeconometrics: Methods and Applications}.
\newblock Cambridge University Press, 2005.

\bibitem{chen2018autoencoder}
Z.~Chen, C.~K. Yeo, B.~S. Lee, and C.~T. Lau.
\newblock Autoencoder-based network anomaly detection.
\newblock In {\em Wireless Telecommunications Symposium (WTS)}, pages 1--5,
  2018.

\bibitem{Coles_2001}
S.~Coles.
\newblock {\em An Introduction to Statistical Modeling of Extreme Values}.
\newblock Springer, 2001.

\bibitem{GIRDLER2021106990}
T.~Girdler and V.~G. Vassilakis.
\newblock Implementing an intrusion detection and prevention system using
  software-defined networking: Defending against {ARP} spoofing attacks and
  blacklisted {MAC} addresses.
\newblock {\em Computers \& Electrical Engineering}, 90:106990, 2021.

\bibitem{Guo_2025}
Z.~Guo, W.~Huang, Y.~Chen, D.~Wang, C.~Gong, and N.~Ma.
\newblock Adaptive multi-scale graph attention and {L}aplacian transform for
  industrial network anomaly detection.
\newblock In {\em IEEE International Conference on Industrial Informatics
  (INDIN)}, 2025.

\bibitem{HernandezCampos2004}
F.~Hern{\'{a}}ndez-Campos, J.~S. Marron, G.~Samorodnitsky, and F.~D. Smith.
\newblock {Variable heavy tails in Internet traffic}.
\newblock {\em Performance Evaluation}, 58(2-3):261--284, 2004.

\bibitem{fpp3}
R.~J. Hyndman and G.~Athanasopoulos.
\newblock {Forecasting: Principles and Practice}.
\newblock 2021.

\bibitem{Kandan_2022}
S.~Kandanaarachchi and R.~J. Hyndman.
\newblock Leave-one-out kernel density estimates for outlier detection.
\newblock {\em Journal of Computational and Graphical Statistics},
  31(2):586--599, 2022.

\bibitem{Kandan_2024}
S.~Kandanaarachchi, C.~Sanderson, and R.~J. Hyndman.
\newblock Extreme value modelling of feature residuals for anomaly detection in
  dynamic graphs.
\newblock In {\em International Conference on Soft Computing and Machine
  Intelligence (ISCMI)}, pages 32--37, 2024.

\bibitem{Ke_2017}
G.~Ke, Q.~Meng, T.~Finley, T.~Wang, W.~Chen, W.~Ma, Q.~Ye, and T.-Y. Liu.
\newblock {LightGBM}: A highly efficient gradient boosting decision tree.
\newblock In {\em Neural Information Processing Systems (NeurIPS)}, 2017.

\bibitem{khamphakdee2014improving}
N.~Khamphakdee, N.~Benjamas, and S.~Saiyod.
\newblock Improving intrusion detection system based on {S}nort rules for
  network probe attack detection.
\newblock In {\em International Conference on Information and Communication
  Technology (ICoICT)}, pages 69--74, 2014.

\bibitem{khraisat2019survey}
A.~Khraisat, I.~Gondal, P.~Vamplew, and J.~Kamruzzaman.
\newblock Survey of intrusion detection systems: techniques, datasets and
  challenges.
\newblock {\em Cybersecurity}, 2(1):1--22, 2019.

\bibitem{Lancewicki_2014}
T.~Lancewicki and M.~Aladjem.
\newblock Multi-target shrinkage estimation for covariance matrices.
\newblock {\em IEEE Transactions on Signal Processing}, 62(24):6380--6390,
  2014.

\bibitem{Ledoit_2022}
O.~Ledoit and M.~Wolf.
\newblock The power of (non-)linear shrinking: A review and guide to covariance
  matrix estimation.
\newblock {\em Journal of Financial Econometrics}, 20(1):187--218, 2022.

\bibitem{provotar2019unsupervised}
O.~I. Provotar, Y.~M. Linder, and M.~M. Veres.
\newblock Unsupervised anomaly detection in time series using {LSTM}-based
  autoencoders.
\newblock In {\em IEEE International Conference on Advanced Trends in
  Information Theory (ATIT)}, pages 513--517, 2019.

\bibitem{Reiss_2007}
R.-D. Reiss and M.~Thomas.
\newblock {\em Statistical Analysis of Extreme Values: with Applications to
  Insurance, Finance, Hydrology and Other Fields}.
\newblock Birkhäuser, 3rd edition, 2007.

\bibitem{Sanderson_2023}
C.~Sanderson, D.~Douglas, and Q.~Lu.
\newblock Implementing responsible {AI}: Tensions and trade-offs between ethics
  aspects.
\newblock In {\em International Joint Conference on Neural Networks (IJCNN)},
  2023.

\bibitem{spiliotis2021hierarchical}
E.~Spiliotis, M.~Abolghasemi, R.~J. Hyndman, F.~Petropoulos, and
  V.~Assimakopoulos.
\newblock Hierarchical forecast reconciliation with machine learning.
\newblock {\em Applied Soft Computing}, 112:107756, 2021.

\bibitem{Sun_2021}
Y.~Sun, H.~Esaki, and H.~Ochiai.
\newblock Adaptive intrusion detection in the networking of large-scale {LANs}
  with segmented federated learning.
\newblock {\em IEEE Open Journal of the Communications Society}, 2:102--112,
  2021.

\bibitem{Talagala2021}
P.~D. Talagala, R.~J. Hyndman, and K.~Smith-Miles.
\newblock {Anomaly Detection in High-Dimensional Data}.
\newblock {\em Journal of Computational and Graphical Statistics},
  30(2):360--374, 2021.

\bibitem{Tariq_2025}
S.~Tariq, M.~Baruwal~Chhetri, S.~Nepal, and C.~Paris.
\newblock Alert fatigue in security operations centres: Research challenges and
  opportunities.
\newblock {\em ACM Computing Surveys}, 57(9), 2025.

\bibitem{Tharwat_2020}
A.~Tharwat.
\newblock Classification assessment methods.
\newblock {\em Applied Computing and Informatics}, 17(1):168--192, 2020.

\bibitem{Wickramasuriya_2019}
S.~L. Wickramasuriya, G.~Athanasopoulos, and R.~J. Hyndman.
\newblock Optimal forecast reconciliation for hierarchical and grouped time
  series through trace minimization.
\newblock {\em Journal of the American Statistical Association},
  114(526):804--819, 2019.

\end{thebibliography}

\end{document}